\documentclass[sigconf]{acmart}
\AtBeginDocument{%
  }
\usepackage{booktabs}
\usepackage{graphicx}
\usepackage{float}
\usepackage{placeins}
\usepackage[table]{xcolor}
\usepackage{soul} 

\usepackage{float}

\usepackage{amssymb}
\usepackage{newtxmath}
\usepackage{multirow} 

\setcopyright{acmlicensed}
\copyrightyear{2025}
\acmYear{2025}
\acmDOI{XXXXXXX.XXXXXXX}
\acmConference[CODS '25]{Make sure to enter the correct
  conference title from your rights confirmation email}{Dec 17-20, 2025}{IISER Pune, Pune, India}

\usepackage{amsmath, amssymb, algorithm, algpseudocode}
\usepackage{algorithmicx}
\usepackage{booktabs}

\definecolor{highlightbold}{RGB}{255, 220, 220}       
\definecolor{highlightunderline}{RGB}{255, 240, 240}  

\begin{document}

\title{Flex-GAD : Flexible Graph Anomaly Detection }

\author{Apu Chakraborty}
\email{apuchakraborty37@gmail.com}
\orcid{0009-0006-0821-3272}
\affiliation{%
  \institution{IIT Bhilai}
  \country{India}
}

\author{Anshul Kumar}
\email{anshulchrs@gmail.com}
\orcid{ 0009-0001-4230-8877}
\affiliation{%
  \institution{IIT Bhilai}
  \country{India}
}

\author{Gagan Raj Gupta}
\email{gagan@iitbhilai.ac.in}
\orcid{0000-0002-8568-2949}
\affiliation{%
  \institution{IIT Bhilai}
  \country{India}
}

\begin{abstract}
Detecting anomalous nodes in attributed networks, where each node is associated with both structural connections and descriptive attributes, is essential for identifying fraud, misinformation, and suspicious behavior in domains such as social networks, academic citation graphs, and e-commerce platforms. We propose Flex-GAD, a novel unsupervised framework for graph anomaly detection at the node level. Flex-GAD integrates two encoders to capture complementary aspects of graph data. The framework incorporates a novel community-based GCN encoder to model intra-community and inter-community information into node embeddings, thereby ensuring structural consistency, along with a standard attribute encoder. These diverse representations are fused using a self-attention-based representation fusion module, which enables adaptive weighting and effective integration of the encoded information. This fusion mechanism allows automatic emphasis of the most relevant node representation across different encoders. We evaluate Flex-GAD on seven real-world attributed graphs with varying sizes, node degrees, and attribute homogeneity. Flex-GAD achieves an \textbf{average AUC improvement of 7.98\%} over the previously best-performing method, GAD-NR, demonstrating its effectiveness and flexibility across diverse graph structures. Moreover, it significantly reduces training time, running \textbf{102$\times$ faster per epoch than Anomaly DAE} and \textbf{3$\times$ faster per epoch than GAD-NR} on average across seven benchmark datasets.
\end{abstract}




\maketitle
\section{Introduction}
Anomaly detection in graph-structured data has become increasingly significant due to the growing reliance on graph-based representations in domains such as social networks, financial systems, and biological networks\cite{dou2020enhancing}\cite{gao2022rumordetectionselfsupervisedlearning}.There are several GAD scenarios, but node-level GAD is particularly important because it focuses on detecting abnormalities in individual nodes. For example, node-level GAD can help identify fraudsters in e-commerce networks, thereby enhancing security and trust within online marketplaces. This paper is specifically focused on tackling the challenging and practical research problem of unsupervised node-level GAD. 

\begin{figure*}[t!]
    \centering
    \includegraphics[width=\textwidth, height=0.42\textheight, keepaspectratio]{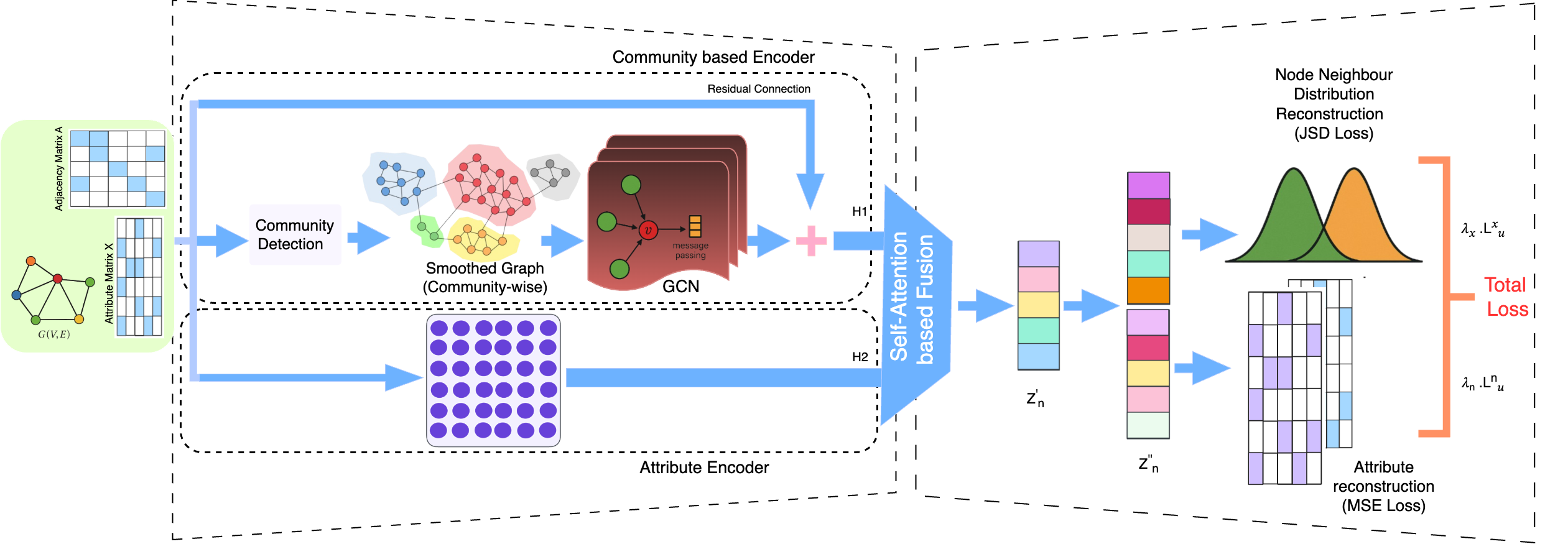}
    \caption{Architecture of the Flex-GAD . The encoder (left) comprises a community-based Graph Convolutional Network (GCN) encoder and a node attribute encoder, whose respective outputs are integrated via a self-attention-based fusion mechanism. The decoder (right) performs two reconstruction tasks: node neighborhood distribution and node attribute reconstruction. }
    \label{fig:model1}
\end{figure*}

Recent approaches leveraging autoencoders and Graph Neural Networks (GNNs)~\cite{kipf2016variationalgraphautoencoders, Roy_2024} have shown promise by encoding both graph structure and node attributes into unified representations for anomaly detection. These methods typically rely on reconstruction loss to distinguish normal data from anomalies, where high reconstruction errors indicate potential anomalies. However, they face several significant challenges that limit their effectiveness:

\begin{enumerate}
    \item \textbf{Representation-collapse:} These models suffer from representation collapse issues~\cite{rusch2023surveyoversmoothinggraphneural, wu2024demystifyingoversmoothingattentionbasedgraph}, where repeated message passing causes node representations to become indistinguishable, reducing the model's discriminative power.
    
    \item \textbf{Imbalance between structure and feature learning:} They often over-emphasize one aspect (either structural or feature-based information) at the expense of the other.
    
    \item \textbf{Poor performance in low-dimensional feature spaces:} Their performance significantly degrades when node attributes provide limited discriminative information.
    
    \item \textbf{Dependence on feature based homophilic assumptions:} These models are heavily dependent on node features and implicitly assume homophilic graph structures, limiting their robustness and applicability in heterophilic networks where dissimilar nodes frequently connect. Since neighboring nodes naturally exhibit different attributes in heterophilic settings, simple attribute based autoencoders cannot distinguish between expected feature dissimilarity and genuine anomalous behavior. For example in a corporate network managers normally communicate with diverse staff in structured ways, anomalies may involve irregular communication bursts to random recipients.
    
    \item \textbf{To much dependency on right hyperparameter:} These methods often require careful tuning of a large number of hyperparameters which is very time consuming, making them difficult to generalize and deploy reliably in practice.
\end{enumerate}

In this paper, we propose \textbf{Flex-GAD (Flexible for Graph Anomaly Detection)} as shown in Figure \ref{fig:model1}, a novel framework designed to address the limitations of existing GAD methods. Flex-GAD incorporates \textbf{community detection} on the adjacency matrix to generate a Smoothed Graph (Community-wise), enhancing the importance of edge features during GCN aggregation and effectively encoding structural information. To the best of our knowledge, the concept of Smoothed Graph (Community-wise) has never been used in graph anomaly detection. The architecture comprises the following key components:

\begin{itemize}
    \item \textbf{Community-Based GCN Encoder}: Captures structural information by leveraging community detection results and aggregates neighborhood features for anomaly detection.
   
    \item \textbf{Attribute Encoder}: Processes node attribute data independently to complement the structual features.
\end{itemize}

The outputs of these two components are combined using a \textbf{representation fusion mechanism} based on self-attention, ensuring that Flex-GAD automatically integrates right amount of structural and attribute-based information for robust anomaly detection. To further improve robustness, Flex-GAD adopts \textbf{Jensen-Shannon Divergence (JSD)} for neighborhood reconstruction during the decoding step, replacing KL divergence to mitigate numerical instabilities caused by overlapping distributions.

Our key contributions are summarized as follows:

\begin{itemize}
    
    \item We introduce a novel approach that incorporates community structures into graph neural network encoders through the concept of \textit{Smoothed Graph (Community-wise)}—a technique not explored in prior work. This enables the effective use of decades of community detection research for structure-based representation learning for unsupervised node graph anomaly detection.

    \item We design a self-attention fusion mechanism that dynamically weights encoder outputs for each node, reducing noise and boosting performance. It consistently outperforms individual encoders across multiple datasets, while its removal leads to significant performance degradation. Additionally, attention scores aid in automated hyperparameter selection during anomaly scoring, removing need of hefty grid search for optimal hyper-parameters.

    \item We adopt Jensen-Shannon Divergence for neighborhood reconstruction, replacing Kullback-Leibler divergence to improve numerical stability when handling overlapping distributions.

    \item Our work highlights the importance of feature-based \\ homophily-heterophily in the context of unsupervised graph anomaly detection.
    
\end{itemize}

Flex-GAD is highly efficient (minimizes average epoch time during training) and adapts effectively to diverse datasets, even in datasets with low dimensional attributes. And when comparing with GAD-NR, the current SOTA GAD-GAE-based approach, our method achieved a 7.79\% improvement in average AUC. Moreover, Flex-GAD significantly reduces training time, running 102× faster per epoch than Anomaly DAE and 3× faster per epoch than GAD-NR on average across seven benchmark datasets. \href{https://github.com/apu20nam/Flex-GAD}{\textbf{(Code available at this link)}}.

\section{Related Work}
Anomaly detection in graph-structured data has become increasingly significant due to wide application in real life scenarios e.g., abusive user behaviors in online user networks, fraudulent activities in financial networks, and spams in social networks. GAD aims to recognize the anomaly instances in graph data that may vary from nodes, edges, to subgraphs by learning an anomaly scoring function.

\textbf{Traditional GAD methods}. They achieve anomaly detection using matrix decomposition and residual analysis like Radar \cite{10.5555/3172077.3172187} analyses the residual between target node attributes and the majority to calculate the anomaly score. ANOMALOUS \cite{10.5555/3304222.3304256} extends the framework of \cite{10.5555/3172077.3172187} by incorporating the CUR decomposition with residual analysis. However, their performance is often bottle necked due to the lack of representation power to capture the rich structure-attribute semantics of the graph data and to handle high-dimensional node attributes. 

\textbf{Graph Neural Networks.} 
Graph neural networks (GNNs) have demonstrated strong capabilities in learning meaningful graph representations by capturing structural patterns. This has led to a wide range of GNN-based approaches for graph anomaly detection (GAD). In particular, deep learning methods such as reconstructive learning have shown significant improvements over traditional approaches.

 DOMINANT~\cite{ding2019deep} is one of the earliest generative GAD methods, learning node embeddings by minimizing reconstruction errors over both attributes and adjacency matrices. AnomalyDAE~\cite{fan2020anomalydae} improves upon this by decoupling attribute and structure encoders for more efficient modeling. Similarly, ComGA~\cite{10.1145/3488560.3498389} integrates community detection within an autoencoder to propagate community-specific information into node representations.

Autoencoder-based techniques assume that normal patterns can be effectively reconstructed, whereas anomalies will yield high reconstruction errors. These methods, widely used in tabular and image domains~\cite{pang2019deepanomalydetectiondeviation}, are equally effective in graph settings. Graph autoencoders (GAEs) extend this idea to graphs by combining structure and attributes through GNNs~\cite{kipf2016variationalgraphautoencoders, xu2019powerfulgraphneuralnetworks, Fan2021ANOMALYDAE}. Reconstruction loss thus becomes a key signal for detecting anomalies.

 Contrastive approaches offer a different perspective. CoLA~\cite{9395172} formulates anomaly detection as a contrastive task, using a discriminator to compare embeddings of target nodes and their neighborhoods. ANEMONE~\cite{zheng2024unsupervisedfewshotgraphanomaly} further extends this by employing patch-level contrastive learning to detect anomalies at multiple scales.

Recent GAD methods also incorporate attention mechanisms. GUIDE~\cite{9671990} uses a graph attention network to focus on the importance of neighbors during reconstruction. GAD-NR~\cite{Roy_2024} aims to reconstruct neighborhood structure and attributes, achieving state-of-the-art results. ComGA enhances structural representation learning by propagating community-level signals through a community-aware encoder based on the graph’s modularity matrix.

 While GAE-based methods have proven effective, they face limitations such as overfitting, high computational cost, and reduced performance with deeper architectures~\cite{zhang2019gresnetgraphresidualnetwork}. Some extensions introduce multi-view reconstruction~\cite{9162509}, enabling richer representation learning from heterogeneous inputs.

A fundamental challenge in GAD is the variation across datasets~\cite{liu2022bondbenchmarkingunsupervisedoutlier}. Some graphs feature dense attributes with sparse connectivity, while others have rich edge structures but limited attribute information. For instance, social networks often contain abundant node features, whereas communication networks may primarily rely on structural signals. These differences highlight the need for adaptive methods that dynamically prioritize either attribute or structure based on the dataset characteristics.

Using community based structures for enhancing  shows promise as seen in LouvainNE 
 a recent development that leverages Louvain-based community detection to construct effective embeddings. Despite its simplicity, it shows competitive performance by exploiting modularity-based structures, and forms a promising direction for integrating classic community detection with modern GNNs.




\section{Preliminaries}
\subsection{Definitions and Notations}

In this section, we introduce the key definitions and notations used throughout the paper. The notations can be found in the notation summary table \ref{tab:notation-summary}.

\begin{definition}[Attributed Graph]
An \emph{attributed graph} is defined as $\mathcal{G} = (\mathcal{V}, \mathcal{E}, \mathbf{X})$, where $\mathcal{V} = \{v_1, v_2, \dots, v_N\}$ denotes the set of nodes with $|\mathcal{V}| = N$, $\mathcal{E}$ denotes the set of edges with $|\mathcal{E}| = E$, and $\mathbf{X} \in \mathbb{R}^{N \times M}$ is the node attribute matrix. The $i$-th row of $\mathbf{X}$, denoted as $\mathbf{x}_i \in \mathbb{R}^M$, corresponds to the attribute vector of node $v_i$. The graph topology is represented by the adjacency matrix $\mathbf{A} \in \{0,1\}^{N \times N}$, where $\mathbf{A}_{ij} = 1$ indicates the existence of an edge between nodes $v_i$ and $v_j$, and $\mathbf{A}_{ij} = 0$ otherwise.
\end{definition}

\begin{definition}[Community]
A \emph{community} refers to a subset of nodes within the graph that are densely connected internally. Each node $v_i \in \mathcal{V}$ is assigned a community label $c_i \in \mathcal{C}$, where $\mathcal{C}$ denotes the set of all communities in the graph.
\end{definition}

\begin{table}[h]
\centering
\begin{tabular}{ll}
\toprule
\textbf{Symbol} & \textbf{Description} \\
\midrule
$\mathcal{G} = (\mathcal{V}, \mathcal{E})$ & Undirected graph with nodes and edges \\
$\mathcal{V}$ & Set of nodes, $|\mathcal{V}| = N$ \\
$\mathcal{E}$ & Set of edges, $|\mathcal{E}| = E$ \\
$\mathbf{X} \in \mathbb{R}^{N \times M}$ & Node attribute matrix \\
$\mathbf{x}_i \in \mathbb{R}^M$ & Attribute vector of node $v_i$ \\
$\mathbf{A} \in \{0,1\}^{N \times N}$ & Adjacency matrix \\
$\mathbf{A}_{ij}$ & Entry indicating edge between $v_i$ and $v_j$ \\
$c_i$ & Community label of node $v_i$ \\
$\mathcal{C}$ & Set of all communities \\
$\mathcal{H}$ & Output of Neural Network \\
\bottomrule
\end{tabular}
\caption{Summary of notations used in the paper.}
\label{tab:notation-summary}
\end{table}

\subsection{Graph Convolutional Networks (GCNs)}

A Graph Convolutional Network (GCN) \cite{kipf2017semisupervisedclassificationgraphconvolutional} is a neural architecture designed to operate on graph-structured data.

To allow information propagation from a node to itself, self-loops are added: $\tilde{A} = A + I_n$, where $I_n$ is the identity matrix. The corresponding degree matrix is $\tilde{D}_{ii} = \sum_j \tilde{A}_{ij}$. 

The layer-wise propagation rule for a GCN is defined as:

\begin{equation}
H^{(l+1)} = \sigma\left( \tilde{D}^{-1/2} \tilde{A} \tilde{D}^{-1/2} H^{(l)} W^{(l)} \right),
\end{equation}

where:
\begin{itemize}
    \item $H^{(l)}$ is the input feature matrix at layer $l$ ($H^{(0)} = X$),
    \item $W^{(l)}$ is a learnable weight matrix at layer $l$,
    \item $\sigma(\cdot)$ is an activation function (e.g., ReLU).
\end{itemize}

This formulation enables each node to update its representation by aggregating and transforming features from its immediate neighbors and itself, normalized by node degrees.

\subsection{Problem Formulation}

In this paper, we focus on the unsupervised node-level graph anomaly detection (GAD) problem. Given an attributed graph G, our model aims to detect nodes
that significantly differ from other nodes from the perspectives of structure and attributes in the graph. We use graph auto encoder (GAE) for unsupervised node-level graph anomaly detection.

\section{Methodology}

In this section, we introduce our Flex-GAD architecture  specified in Figure~\ref{fig:model1}. The Flex-GAD architecture employs two encoders: the first encoder utilises a modified Graph Convolution Network (GCN) and the second encoder is an attribute encoder for handling node-specific attributes\cite{corso2020principal,corsini2021selfsupervised}. After that we used our novel self attention based representation fusion technique to fuse the two representations. Then we used a simple concatenation technique and passed this final embedding as an input of decoder for  attribute reconstruction and node neighbor distribution reconstruction. Our approach is elaborated in Algorithm \ref{algo}.

\subsection{Encoder Architecture}

\begin{figure}[h!]
    \centering
    \includegraphics[scale=0.05]{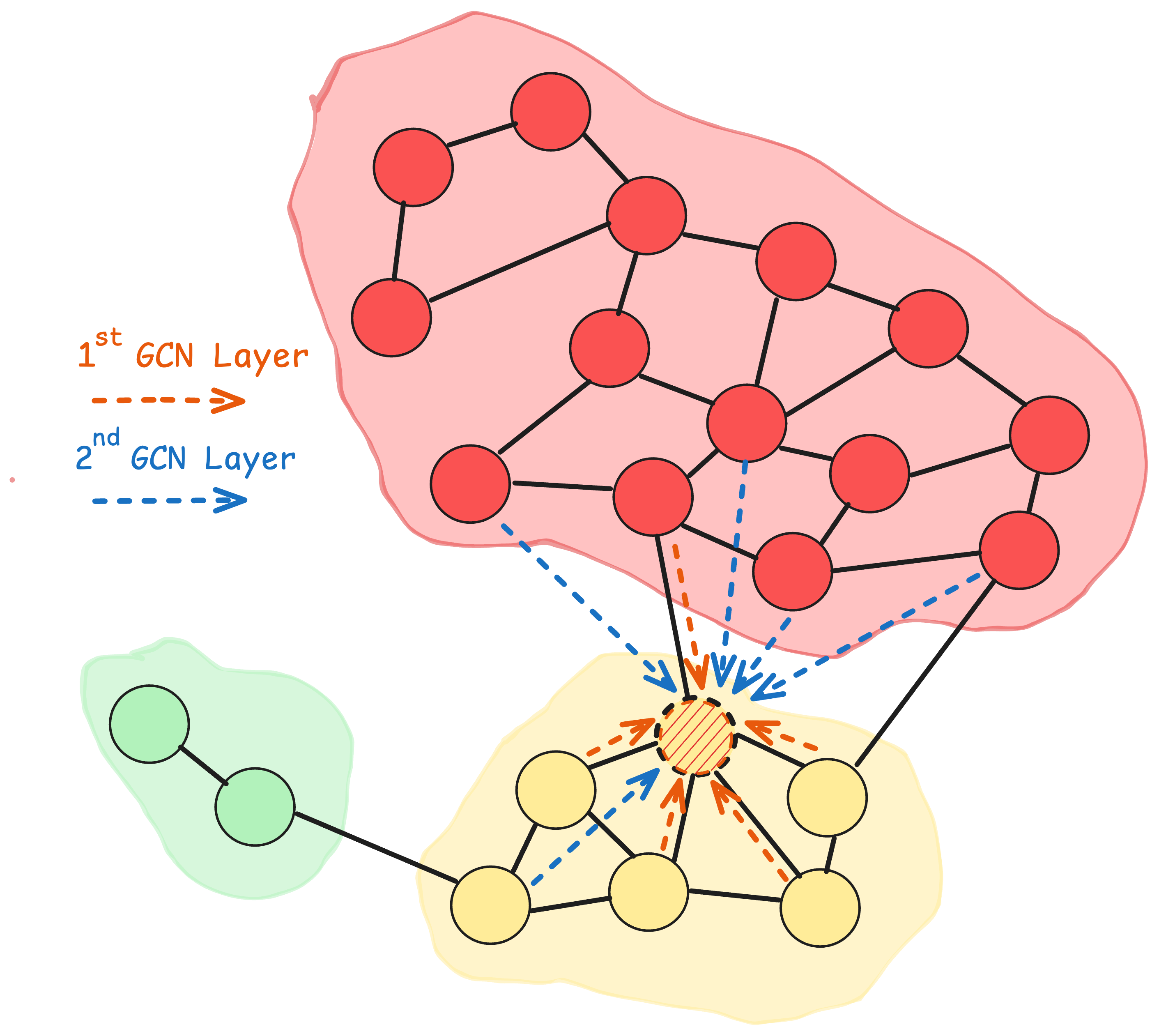}
    \caption{Community-wise smoothed graph with 2-layer GCN updates }
    \label{fig:dummy_GCN}
\end{figure} 
\subsubsection{\textbf{Community Based Encoder :}}
Flex-GAD utilized  a community-based Graph Convolutional Network (GCN) encoder that integrates community structure into the node representation learning process. The encoder is designed to learn structural properties of the graph while addressing the over-smoothing problem common in deep GCNs. The approach consists of three main components: community-averaged initialization, structural message passing via GCN layers, and feature preservation through residual connections .

Given an input graph, we first apply a community detection algorithm to identify node groupings. Communities in graphs represent groups of nodes that are more densely connected internally than to the rest of the graph, forming hierarchical structural units that exhibit local consistency in connectivity patterns while maintaining distinct boundaries from other communities. Each node is then initialized with a feature vector that is the average of features of all nodes within its community. This initialization step standardizes node representations within communities, removing reliance on potentially noisy individual features and focusing the Felx-GAD's learning capacity on structural differences and inter-community relations.
\[
x_i^{\text{new}} = \text{avg}(x_j \mid c_j = c_i)
\]

After initialization, we perform message passing using a multi-layer GCN architecture over the Smoothed Graph (Community-wise) as shown in Figure \ref{fig:dummy_GCN} . As nodes within the same community begin with identical features, their representations evolve differently depending on their local structure—particularly the number of connections and the communities their neighbors belong to. This process allows the Flex-GAD to encode neighborhood distribution patterns, capturing both intra-community consistency and inter-community variation.

Multi-hop message passing further enables the encoder to integrate higher-order structural signals. Nodes that serve as bridges between communities receive more diverse inputs, resulting in more distinct embeddings. This mechanism helps the Flex-GAD to learn modular patterns and connectivity differences across the graph.

Now to add this structural information back to node embeddings we use a residual connections that add the original node feature features back to the output of the final GCN layer. This help retain identity of each node while enriching it with structural information accumulated through message passing.

By combining these elements, our Flex-GAD introduces a structure-aware encoding scheme that emphasizes relational and topological features of the graph while preserving meaningful feature diversity across nodes.

\subsubsection{\textbf{ Attribute Encoder}}
In addition to the structural encoder, Flex-GAD employs an attribute encoder for the latent attribute embedding \(Z^A\). The attribute encoder consists of non-linear feature transformation layers that map observed attribute data into a latent space\cite{Fan2021ANOMALYDAE}\cite{fan2020anomalydae}. The transformation is given by:
\[Z^a = \alpha(W \cdot X + b)\]
 \[
    H_2 = Z^a
    \]
where \(W \in \mathbb{R}^{d \times m}\), \(X \in \mathbb{R}^{m}\), and \(b \in \mathbb{R}^{d}\). Here, \(m\) is the input dimension, and \(d\) is the hidden dimension.

\subsection{Self Attention Based Representation Fusion}

Flex-GAD employes a novel implementation of self attention for representation fusion to combine the final embedding representations adaptively from the two encoders. Fusing multiple graph representations is a non-trivial task. Multi-representation fusion refers to strategies that combine information from multiple data modalities to enhance learning.

Although attention mechanisms have previously been applied in graph neural networks, such as in the Graph Attention Network (GAT), they have not been employed between representations of the same nodes from different encoders. By providing an adaptive mechanism that responds to the nature of the graph, this approach enables Flex-GAD to choose between the structural information from a community-based GCN encoder and the feature information from a node attribute encoder.

Flex-GAD first applied self attention~\cite{vaswani2023attentionneed} to the two final embeddings to enable interaction between them. This mechanism facilitates information exchange between the representations and allows the Flex-GAD to adapt to the dataset by determining the relative importance of structural, feature-based representation or a mixture of both the representation.

Let \( H_1 \) and \( H_2 \) be the final embeddings from the two encoders. We compute the attention-enhanced representations as follows:
\[
H_1', H_2' = \text{self\_attn}(H_1, H_2)
\]

We then concatenate these enhanced embeddings:
\[
z_n = \text{concat}(H_1', H_2'), \quad z_n \in \mathbb{R}^{n \times 2d}
\]

Finally, we apply a linear transformation without bias:
\[
z_n' = z_n W_2, \quad W_2 \in \mathbb{R}^{2d \times d}, \quad z_n' \in \mathbb{R}^{n \times d}
\]

\subsection{Decoder Architecture}
Flex-GAD employs two decoder, for the reconstruction process. The one decoder focuses on \textbf{Node Neighborhood Distribution Reconstruction } and other on \textbf{Attribute Reconstruction}.

\subsubsection{\textbf{Attribute-Reconstruction}}
For each node \(u\), the self-recon structed representation \(\hat{h}_u^{(0)}\) is derived from the input node representation \(Z^n\) using a transformation function \(\Phi_x\):
\[\hat{h}_u^{(0)} = \Phi_x(Z^{(n)})\]
The self-reconstruction loss, which measures the discrepancy between the original and reconstructed node representations, is computed using a distance function \(\mathcal{D}(\cdot)\), typically the L2-distance:
\[\mathcal{L}_u^x = \mathcal{D}(h_u^{(0)}, \hat{h}_u^{(0)})\]


\subsubsection{\textbf{Node Neighbor  Distribution Reconstruction}}
To reconstruct the distribution \( P_u \) from the node representation \( \mathbf{h}_u^{(L)} \), we first map \( \mathbf{h}_u^{(L)} \) to an estimated distribution \( \hat{P}_u \).

Computing the divergence between the true distribution \( P_u \) and the estimated distribution \( \hat{P}_u \) is a critical step in guiding the model to learn meaningful representations. While previous approaches have commonly used the Kullback--Leibler (KL) divergence for this purpose, it can be numerically unstable, especially when the distributions have non-overlapping support.

Inspired by GAD-NR, we approximate both \( P_u \) and \( \hat{P}_u \) as multivariate Gaussian distributions:
\[
P_u \sim \mathcal{N}(\boldsymbol{\mu}_u, \boldsymbol{\Sigma}_u), \quad
\hat{P}_u \sim \mathcal{N}(\hat{\boldsymbol{\mu}}_u, \hat{\boldsymbol{\Sigma}}_u),
\]
where \( \boldsymbol{\mu}_u, \hat{\boldsymbol{\mu}}_u \) denote the mean vectors, and \( \boldsymbol{\Sigma}_u, \hat{\boldsymbol{\Sigma}}_u \) denote the covariance matrices.

Instead of using the KL divergence, we propose using the Jensen--Shannon divergence (JSD) as the reconstruction loss:
\[
\mathcal{L}_{\text{neigh}}(u) = \mathrm{JSD}(P_u \| \hat{P}_u),
\]
where JSD is defined as
\[
\mathrm{JSD}(P \| Q) = \frac{1}{2} \mathrm{KL}\left(P \Big\| \frac{P + Q}{2}\right) + \frac{1}{2} \mathrm{KL}\left(Q \Big\| \frac{P + Q}{2}\right).
\]

The Jensen--Shannon divergence is symmetric and more numerically stable than KL divergence, making it a suitable choice for measuring the divergence between the estimated and target distributions in our framework.

\subsubsection{\textbf{Total Loss}} \label{sec:total_loss}
The total loss for is the weighted summation between the loss during node neighbhorhood distribution recostruction and self reconstruction:
\[\mathcal{L} = \sum_{u \in V} \left( \lambda_x \mathcal{L}_u^x  + \lambda_n \mathcal{L}_u^n \right)\]

\subsubsection{\textbf{Anomaly Detection}} \label{sec:score}

We calculate the composite anomaly score by summing two normalized per-node loss components: self-reconstruction loss and neighbor distribution reconstruction loss. The final anomaly score for each node is computed using a weighted sum of the two losses:

\begin{align*}
\text{Score}^{(i)} = \lambda'_n \cdot \text{h\_loss}^{(i)} + \lambda'_x \cdot \text{feature\_loss}^{(i)}
\end{align*}

In the self-attention mechanism, the output of each encoder is modulated by attention scores relative to the other encoder. These attention scores can be leveraged to derive optimal hyperparameters. Specifically, self-attention is applied between representations of the same node across different encoder variants. We average the attention scores across all nodes, resulting in a final \(2 \times 2\) attention matrix, as shown below:

\begin{center}
\begin{tabular}{c|cc}
           & \textbf{Encoder 1} & \textbf{Encoder 2} \\
\hline
\textbf{Encoder 1} & a & b \\
\textbf{Encoder 2} & c & d \\
\end{tabular}
\end{center}

We use this matrix to determine the weights for the loss components:

\begin{align*}
\lambda'_n \cdot \text{h\_loss}^{(i)} = a + c \\
\lambda'_x \cdot \text{feature\_loss}^{(i)} = b + d
\end{align*}

The intuition is that the cumulative attention received by each encoder during training reflects its relative importance. For instance, the community-based GCN encoder (Encoder 1), which captures neighborhood structure, contributes to \(\text{h\_loss}\), while the attribute encoder (Encoder 2) contributes to \(\text{feature\_loss}\). The total attention weight received by each encoder is thus used to determine the weight of its corresponding loss term in the final anomaly score.

This mechanism enables automated hyperparameter tuning based on the relative importance of structural and contextual information as learned during training.

Instead of using a fixed threshold for classification, anomaly detection is conducted in a ranking-based manner using the AUC metric, which evaluates the model’s ability to rank anomalous nodes higher than normal ones.

\begin{algorithm}
\caption{Flex-GAD: Flexible Graph Anomaly Detection}
\begin{algorithmic}[1]
\Require Graph $G(V, E, X)$
\Ensure Multi-modality representation and anomaly scores

\State \textbf{Community-based Feature Transformation:}
\[
x_{new,i} = \frac{1}{|C_i|} \sum_{j \in C_i} x_j
\]
\State $X_{new} \gets$ community-averaged features

\State \textbf{Encoder 1 (Community-based encoder):}
\For{$u \in V$}
    \State $h_u^{(0)} \gets \xi(x_{new,u})$
    \State $h_u^{(1)} \gets \text{UPDATE}\left(h_u^{(0)}, \text{AGG}\{ h_v^{(0)} : v \in N_u \}\right)$
\EndFor
\State $H_1 \gets \{ h_u^{(1)} + \hat{A}XW_{\text{residual}} \}_{u \in V}$

\State \textbf{Encoder 2 (Attribute encoder):}
\[
H_2 = \alpha(WX + b)
\]

\State \textbf{Multi-modality Fusion:}
\[
H'_1, H'_2 = \text{SelfAttention}(H_1, H_2)
\]
\State Concatenate the outputs:
    \[
    z_n = \text{concat}(H'_1, H'_2), \quad z_n \in \mathbb{R}^{n \times 2d}
    \]
\State Project to a lower dimension:
    \[
    z'_n = z_n W_1, \quad W_1 \in \mathbb{R}^{2d \times d}, \quad z'_n \in \mathbb{R}^{n \times d}
    \]
\State Project to a higher dimension :
    \[
    z''_n = z'_n W_2, \quad W_2 \in \mathbb{R}^{d \times 2d}, \quad z''_n \in \mathbb{R}^{n \times 2d}
    \]
    \[ Z_n=z''_n
    \]
\State $H''_1, H''_2 \gets \text{Deconcat}(Z'')$

\State \textbf{Decoder:}
\State Reconstruct features: $\hat{x}_u = \Phi_x(H''_{2,u})$
\State Reconstruct neighborhood distribution:
\[
\begin{aligned}
\mu_{\text{true}} &= \text{MeanAgg}\left(\{h_u | u \in N(v)\}\right), \quad \sigma_{\text{true}} = \text{StdAgg}(\cdot) \\
\mu_{\text{gen}} &= \text{MLP}_{\mu}(H''_{1}), \quad \sigma_{\text{gen}} = \exp(\text{MLP}_\sigma(H''_{1}))
\end{aligned}
\]

\State \textbf{Loss:}
\[
\mathcal{L} = \lambda_x \sum_{u \in V} D(x_u, \hat{x}_u) + \lambda_n \mathbb{E}_{v \in V}[\text{JSD}(P_v^{\text{true}} \parallel P_v^{\text{gen}})]
\]

\State \textbf{Anomaly Score:}
\[
\text{Score}_u = \lambda'_x \cdot \text{feature\_loss}_u + \lambda'_n \cdot \text{neigh\_loss}_u
\]
\State Rank nodes by score for anomaly detection using AUC.
\end{algorithmic}
\label{algo}
\end{algorithm}

\section{Datasets} \label{sec:dataset}

We incorporate seven real-world datasets—Cora, Weibo, Reddit, Disney, Books, Enron and  Amazon as summarized in Table~\ref{tab:datasets}.These datasets represent diverse graph types, which are essential for evaluating the robustness of anomaly detection models. Traditional homophily measures rely on labels to compute the fraction of edges connecting nodes within the same class. However, since anomaly detection is often approached using unsupervised techniques, we propose a label-independent approach using feature similarity, thus we calculate homophily on the basis of feature-based similarity between connected nodes. Specifically, we compute a continuous measure of homophily based on the cosine similarity of feature vectors between connected nodes:

\begin{equation} H = \frac{1}{|E|} \sum_{(u,v) \in E} \frac{\mathbf{x}_u \cdot \mathbf{x}_v}{|\mathbf{x}_u| |\mathbf{x}_v|} \end{equation}

where $\mathbf{x}_u$ and $\mathbf{x}_v$ are the feature vectors of nodes $u$ and $v$, and $E$ is the set of edges. This formulation provides a continuous, label-free measure of homophily, enabling consistent comparisons across datasets. Table~\ref{tab:datasets} provides detailed overview of the datasets including the approximation of the \textit{Homophily Ratio} using the above described method.

\begin{table}[!htbp]
\centering
\renewcommand{\arraystretch}{1.2}
\setlength{\tabcolsep}{1.7pt} 
\resizebox{\columnwidth}{!}{ 
\begin{tabular}{lccccccc}
\toprule
\textbf{Dataset} & \textbf{\#Nodes} & \textbf{\#Edges} & \textbf{\#Feat.} & \textbf{Avg. Degree} & \multicolumn{2}{c}{\textbf{Ratio}} \\ 
\cmidrule(lr){6-7}
& & & & & \textbf{Anomaly} & \textbf{Homophily} \\ 
\midrule
Cora & 2,708 & 11,060 & 1,433 & 4.1 & 5.1\% & 0.15 \\ 
Weibo & 8,405 & 407,963 & 400 & 48.5 & 10.3\% & 0.995 \\ 
Reddit & 10,984 & 168,016 & 64 & 15.3 & 3.3\% & 0.978 \\ 
Disney & 124 & 335 & 28 & 2.7 & 4.8\% & 0.763 \\ 
Books & 1,418 & 3,695 & 21 & 2.6 & 2.0\% & 0.670 \\ 
Enron & 13,533 & 176,987 & 18 & 13.1 & 0.4\% & 0.653 \\ 
Amazon & 13,752 & 515,042 & 767 & 37.2 & 5.0\% & 0.798 \\ 
\bottomrule
\end{tabular}
}
\caption{Summary of the seven real-world datasets.}
\label{tab:datasets}
\end{table}

The homophily ratios reveal significant variations across datasets. For instance, Reddit exhibits near-perfect feature based homophily (0.978) , suggesting strong feature alignment among connected nodes, while Cora shows minimal  feature based homophily (0.15), though Cora according to labels is considered a homophily graph of paper citations.

These differences have direct implications:

\textbf{High Homophily} (e.g. Reddit, Disney): Nodes with low feature similarity to their neighbors are likely anomalies in a homophilly setup, making feature learning more important and feature reconstruction methods effective for anomaly detection.

\textbf{Low Homophily} (e.g., Cora): Structural patterns may dominate over feature relationships, favoring topology-aware learning, making neighborhood reconstruction methods effective for anomaly detection.

This analysis highlights the necessity for adaptive anomaly detection frameworks that can dynamically adjust to varying datasets. Such frameworks should be capable of leveraging feature information, structural patterns, or both, depending on the dataset's characteristics. The continuous feature based homophily metric provides a means to assess the dataset’s inherent connectivity patterns, enabling practitioners to tailor detection methods accordingly. 

\section{Experimental Setting}
\subsection{Experimental Goals}
\subsubsection{\textbf{Q1) Performance Comparison:}} How does the proposed method compare to the existing methods across diverse graphs?
\subsubsection{\textbf{Q2) Independent relevance of each encoder and role of Self attention based fusion:}} How does each encoder perform independently and does the whole architecture help improve overall performance ?

\subsubsection{\textbf{Q3) Role of chosen community detection algorithm in the proposed method:}} Can the proposed method perform consistently across all diverse datasets with any community detection algorithm ?

\subsection{Experimental Setup}
In this paper, we adopted the same experimental setup as outlined in the benchmark paper GAD-NR\cite{Roy_2024} and the outlier node detection (BOND) paper\cite{liu2022bondbenchmarkingunsupervisedoutlier}. The datasets used include Weibo, Reddit, Disney, Books, Enron, cora and amazon ; all containing real-world anomaly labels . We reported the mean AUC score, standard deviation, and the best AUC score achieved over 10 runs of the experiment on the entire dataset. AUC score measures the effectiveness of the model's ability to detect anomalies. This is consistent with the methodology used in the benchmark paper~\cite{Roy_2024, liu2022bondbenchmarkingunsupervisedoutlier}. This ensures a fair comparison with baseline approaches.

\textbf{Hyperparameter Tuning:}
We conducted a grid search for the hyper-parameters of our model on each dataset as follows:

\begin{itemize}
    \item $\lambda_x \in \{0.1, 0.3, 0.5, 0.7, 0.8, 0.9, 10, 20, 70, 300, 1000, 3000\}$
    \item $\lambda_n \in \{0.0001, 0.001, 0.5, 0.8, 0.9, 1.0\}$
    \item $d \in \{8, 16, 32, 64, 128, 256\}$
\end{itemize}

From the different combinations tested, as given above, we identified the best configuration for each dataset on the basis of loss convergence and best AUC, which led to the optimal results for our Flex-GAD model as shown in Table \ref{tab:dataset_params}.

\begin{table}[h!]
\centering
\begin{tabular}{lccc}
\toprule
\textbf{Dataset} & $\lambda_x$ & $\lambda_n$ & $d$ \\
\midrule
Cora   & 0.9 & 0.25 & 128 \\
Books  & 0.7 & 0.1  & 8  \\
Disney & $0.7\times10^2$ & 0.1 & 8  \\
Weibo  & $0.1\times10^4$ & 0.5 & 256  \\
Reddit & $0.2\times10^3$ & 0.5 & 32  \\
Enron  & $0.1\times10^2$ & 0.1 & 8 \\
Amazon & 0.4 & 0.1 & 256  \\
\bottomrule
\end{tabular}
\caption{Optimal hyperparameters for Flex-GAD and hidden dimensions used for different datasets.}
\label{tab:dataset_params}
\end{table}

\textbf{Hardware:}All the experiments are performed on a Linux server with a 1800MHz and 1 NVIDIA RTX A6000 GPU with 48GB memory

\section{Results and Discussion }
\subsection{Q1) Performance Comparison:}
\FloatBarrier
\begin{table*}[!htbp]
\centering
\renewcommand{\arraystretch}{1.2}
\setlength{\tabcolsep}{4pt} 
\resizebox{\textwidth}{!}{ 
\begin{tabular}{lccccccccc}
\toprule
\textbf{Algorithm }  & \textbf{Books} & \textbf{Enron} & \textbf{Amazon} &    \textbf{Cora} & \textbf{Weibo} & \textbf{Reddit} & \textbf{Disney} & \textbf{Avg.}\\ 
\midrule
IF\cite{10.1145/2133360.2133363}  & 43.0 ± 1.8  & 40.1 ± 1.4  & 55.2±0.0   & 64.4 ± 1.5  & 53.5 ± 2.8  & 45.2 ± 1.7  & 57.6 ± 2.9 & 50.07 \\ 
LOF\cite{10.1145/342009.335388}  & 36.5 ± 0.0  & 46.4 ± 0.0  & 51.3±3.0    & 69.9 ± 0.0  & 56.5 ± 0.0  & 57.2 ± 0.0  & 47.9 ± 0.0 & 53.94 \\ 
DOMINANT\cite{doi:10.1137/1.9781611975673.67} & 50.1 ± 5.0  & 73.1 ± 8.9  & 81.3±1.0   & 82.7 ± 5.6  & 85.0 ± 14.6  &{56.0 ± 0.2 } & 47.1 ± 4.5 & 62.55  \\ 
CONAD\cite{10.1007/978-3-031-05936-0_35}  & 52.2 ± 6.9  & 71.9 ± 4.9  &   80.5±4.0  & 78.8 ± 9.6  & 85.4 ± 14.3  & 56.1 ± 0.1  & 48.0 ± 3.5 & 61.06\\ 
SCAN\cite{10.1145/1281192.1281280}  & 49.8 ± 1.7  & 52.8 ± 3.4  & 62.2±4.9 & 62.8 ± 4.5  & 63.7 ± 5.6  & 49.9 ± 0.3  & 50.5 ± 4.0 & 58.55\\ 
DONE\cite{10.1145/3336191.3371788}  & 43.2 ± 4.0  & 46.7 ± 6.1  &  82.8±8.8   &{82.4 ± 5.6 } & 85.3 ± 4.1  & 53.9 ± 2.9  & 41.7 ± 6.2 & 64.01\\ 
MLPAE\cite{inproceedings}  & 42.5 ± 5.6 & 73.1 ± 0.0  & 74.2±0.0   & 70.9 ± 0.0  & 82.1 ± 3.6  & 50.6 ± 0.0  & 49.2 ± 5.7 & 63.47 \\ 
Radar\cite{10.5555/3172077.3172187}  & 52.8 ± 0.0  & 80.8 ± 0.0  & 71.8±1.1 & 65.0 ± 1.3  & \cellcolor{highlightbold}\textbf{98.9 ± 0.1 } & 54.9 ± 1.2  & 51.8 ± 0.0 & 54.17 \\ 
AdONE\cite{10.1145/3336191.3371788}  & 53.6 ± 2.0  & 44.5 ± 2.9  &\cellcolor{highlightbold}\textbf{86.6±5.6}  & 81.5 ± 4.5  & 84.6 ± 2.2  & 50.4 ± 4.5  & 48.8 ± 5.1 & 66.71\\ 
ANOMALOUS\cite{10.5555/3304222.3304256}  & 52.8 ± 0.0  & 80.8 ± 0.0  & 72.5±1.5 & 55.0 ± 10.3  & \cellcolor{highlightbold}\textbf{98.9 ± 0.1 } & 54.9 ± 5.6  & 51.8 ± 0.0 & 55.63 \\ 
GUIDE\cite{9671990}  & 48.4 ± 4.6  & OOM\_C  & OOM\_C& 74.7 ± 1.3  & OOM\_C & OOM\_C & 38.8 ± 8.9 & 48.37 \\ 
GCNAE\cite{kipf2016variationalgraphautoencoders}  & 50.0 ± 4.5  & 66.6 ± 7.8  & 74.2±0.0   & 70.9 ± 0.0  & 90.8 ± 1.2  & 50.6 ± 0.0  & 42.2 ± 7.9 & 66.48 \\ 
GAAN\cite{10.1145/3340531.3412070} & 54.9 ± 5.0  & 73.1 ± 0.0  & 80.8±0.3 & 74.2 ± 0.9  &\cellcolor{highlightunderline}\underline{92.5 ± 0.0 } & 55.4 ± 0.4  & 48.0 ± 0.0  & 69.32 \\ 
AnomalyDAE\cite{fan2020anomalydae} & 62.2 ± 8.1  & 54.3 ± 11.2  & 85.7±2.9  &{83.4 ± 2.3 } & 91.5 ± 1.2  & 55.7 ± 0.4  & 48.8 ± 2.2 & 70.55\\ 
GAD-NR\cite{Roy_2024}  & \cellcolor{highlightunderline}\underline{65.7 ± 4.9 } & \cellcolor{highlightunderline}\underline{80.8 ± 2.9 } & 63.76±2.3 & \cellcolor{highlightunderline}\underline{87.5 ± 2.5 } & {87.7 ± 5.3 } &\cellcolor{highlightunderline}\underline {57.9 ± 1.6 } & \cellcolor{highlightunderline}\underline{76.7 ± 2.7 } & 74.19 \\ 
\textbf{Flex-GAD }  & \cellcolor{highlightbold}\textbf{68.8 ± 1.35} & \cellcolor{highlightbold}\textbf{83.21 ± 0.77 } & \cellcolor{highlightunderline}\underline{85.07±2.81}  & \cellcolor{highlightbold}\textbf{92.07±2.89} & {91.50±0.21} & \cellcolor{highlightbold}\textbf{60.07±1.12} & \cellcolor{highlightbold}\textbf{83.67±3.49} & \cellcolor{highlightbold}\textbf{80.63} \\ 
\bottomrule
\end{tabular}
}
\caption{Performance comparison of various algorithms across different datasets. The results are presented for each dataset, showing the mean AUC score ± standard deviation over 10 runs, with the best score achieved in parentheses. \texttt{OOM\_C} indicates the algorithm ran out of memory on the corresponding dataset. Best performances are highlighted in bold, while the second-best performances are underlined.}
\label{table:Results}
\end{table*}
In Table~\ref{table:Results}, we present the results of Flex-GAD on 7 anomaly detection benchmark datasets and compare them with 15 baseline models. Here we used louvian community detection algorithm for out community based GCN. We compare the average AUC score for each dataset. Also, we present the average of the average AUC scores across all 7 datasets in the last column of the table. We achieved the state-of-the-art performance on 5 out of 7 datasets and the highest overall average score as shown in Table~\ref{table:Results}.

For Amazon, the top-performing algorithm is AdONE~\cite{10.1145/3336191.3371788} but it is computationally (167x) higher per epoch as compared to our algorithm. For Weibo, Radar~\cite{10.5555/3172077.3172187} and ANOMALOUS~\cite{10.5555/3304222.3304256} deliver the best results with AUCs of 98.9, by filtering noisy attributes and it is computationally (24x) higher per epoch as compared to our algorithm. It is important to note that Radar and ANOMALOUS are not an autoencoder-based method and therefore is not directly comparable within the scope of encoder-based approaches. Flex-GAD is also very close in performance to AdONE with very less standard deviation compared to it.

Flex-GAD consistently outperforms all other methods across the remaining five datasets: Reddit, Disney, Books, Enron and Cora. As shown in Table~\ref{tab:datasets}, these datasets vary significantly in the number of nodes, edges, feature dimension, feature based homophily, and anomaly ratio. Our architecture of one encoder for structural information learning and other for node feature information learning with a self-attention based fusion for automated adaptation proves to be highly effective and robust in enhancing anomaly detection performance. It achieves an overall average AUC score of 80.63 across the seven benchmarks as shown in the last column of Table~\ref{table:Results}, providing an overall improvement of 7.98\% over GAD-NR. Given that none of the anomaly detection techniques achieves the best performance on all datasets, we conclude that Flex-GAD is the new state-of-the-art.

Our model significantly reduces running time during training: (102x) per epoch compared to the Anomaly DAE and (3x) per epoch as compared to GAD-NR on average across 7 benchmarks, as shown in the Table \ref{table:performance_comparison}. This comparison does not account for GAD-NR's reliance on an extensive grid search to pre-select the best encoder. In contrast, our method incorporates a fusion-based technique that identifies the most suitable encoder features during training, thus eliminating the need for an expensive pre-processing phase.

\begin{table}[ht]
\centering
\renewcommand{\arraystretch}{1.5}
\setlength{\tabcolsep}{1.3pt} 
\renewcommand{\arraystretch}{1} 
\begin{tabular}{lcccccccc}
\toprule
 \textbf{Algorithm} & \textbf{Reddit} & \textbf{Weibo} & \textbf{Books} & \textbf{Cora} & \textbf{Enron} & \textbf{Disney} &\textbf{Amazon}  \\ 
\midrule
DONE & 36 & 25.45 & 3.95 & 0.51 & 50.25 & 4.95 & 56 \\
AdONE& 36 & 26 & 8.15 & 8.12 & 47.8 & 7.8 & 54.18 \\
Anomalous & 1.35 & 2.5 & 0.6 & 2.26 & 1.8 & 0.3 & 17.48  \\
AnomalyDAE & 33 & 24 & 2.8 & 5.6 & 48.57 & 4.15 & 0.91 \\ 
GAD-NR & 0.1 & 0.201 & 0.09 & 2.55 & 0.05 & 0.13 & 0.15 \\  
Flex-GAD & 0.16 & 0.32 & 0.06 & 0.12 & 0.1 & 0.06 & 0.34  \\
\bottomrule
\end{tabular}
\caption{Running time per epoch (in seconds) during training. Comparison between Flex-GAD and important baselines}
\label{table:performance_comparison}
\end{table}

\subsection{Q2) Independent relevance of each encoder and role of Self attention based fusion:}

\begin{figure}[t]
    \centering
  \includegraphics[width=2\textwidth, height=0.28\textheight, keepaspectratio]{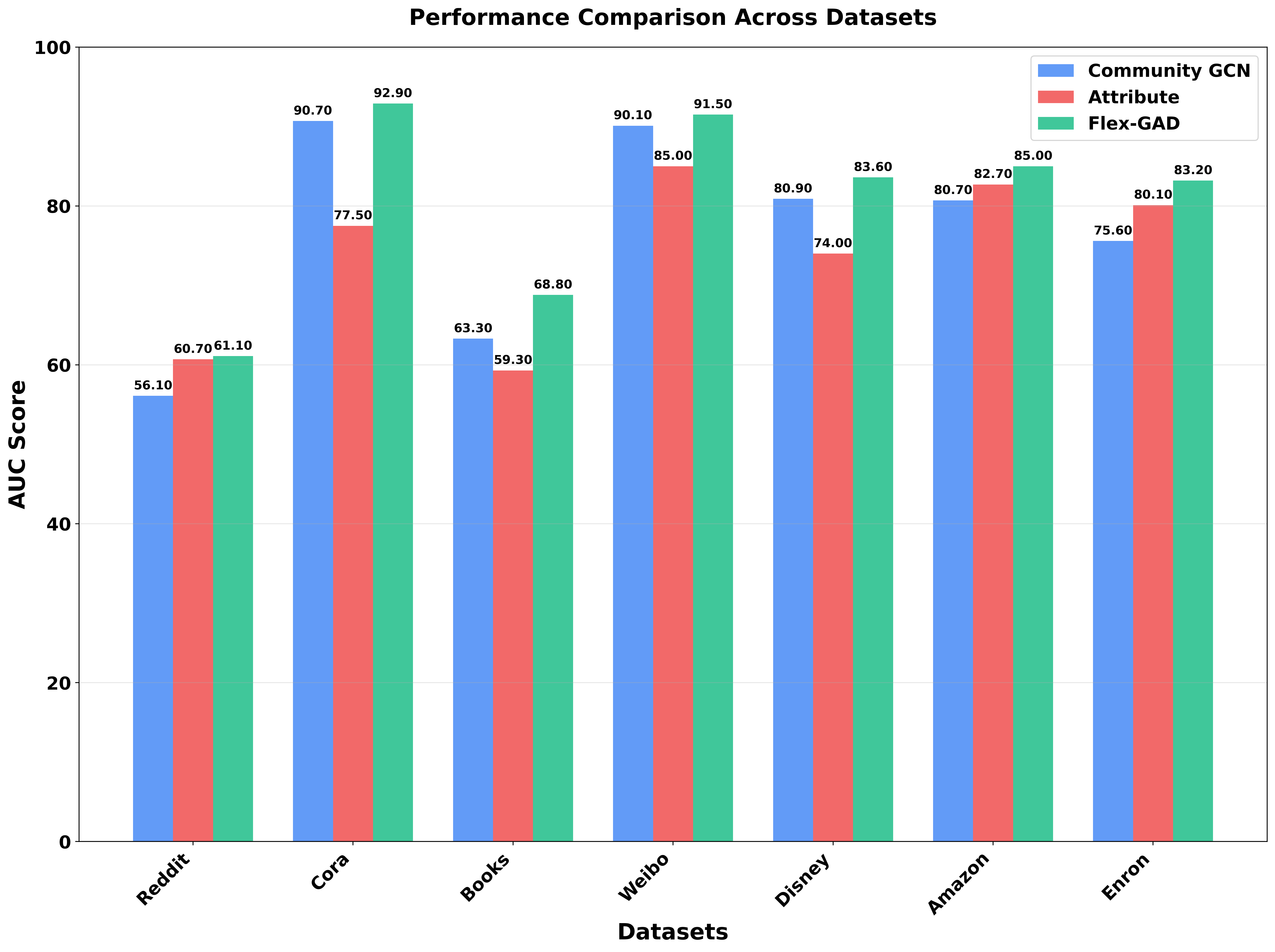}
    \caption{Ablation experiment to show the importance of both encoders and self attention in Flex-GAD.   }
    \label{fig:ablation}
\end{figure}

\begin{figure}[t]
    \centering
  \includegraphics[width=1.8\textwidth, height=0.22\textheight, keepaspectratio]{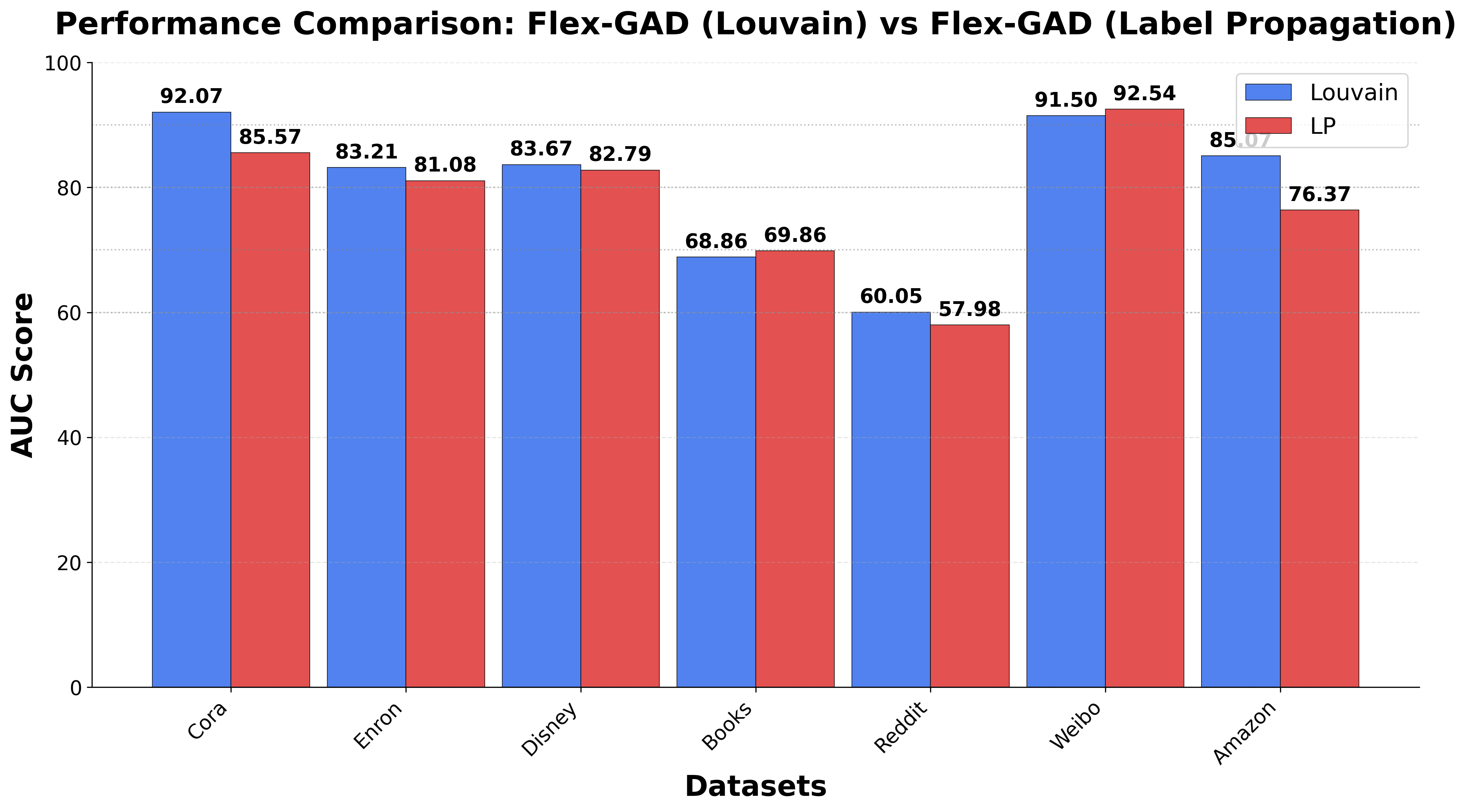}
    \caption{Flex-GAD using different community detection algorithm }
    \label{fig:diff_comm}
\end{figure}

To better understand the contributions of different components in Flex-GAD, we conduct an ablation study.

Flex-GAD integrates two encoders: a community-based GCN (Encoder 1) for structural learning,  attribute encoder (Encoder 2) for capturing node attributes. We can see in Figure~\ref{fig:ablation} that no single encoder performs best in all datasets, our self-attention fusion module selects and combines these encoder representations adaptively, as part of an optimization objective designed to minimize loss Section\ref{sec:total_loss}. This ensures that the most informative representation—or a mixture thereof—is propagated . In several datasets, our self-attention-based fusion module outperforms all individual encoders, and removing it leads to a drop in performance. 

\subsection{Q3) Role of chosen community detection algorithm in the proposed method:}

We compared the performance of Flex-GAD using two famous community detection algorithm, Louvian community detection and label propagation algorithm, as seen in Figure~\ref{fig:diff_comm}. It can be observed that there isn't a very drastic drop in performance. Some variation can be seen due to the varying ability of the community detection algorithm to work in graphs of different sizes. Here modularity high modularity is preferred. Louvian community detection can be seen as a good fit since it optimizes modularity during the community detection process.

\section{Conclusion}
In this paper, we successfully designed an efficient and flexible autoencoder-based unsupervised node anomaly detection method, Flex-GAD. Flex-GAD proposed several novel features to this line of research: incorporation of community structures via Smoothed Graph, self-attention fusion mechanism, automated hyper-parameter selection and JSD loss during neighborhood reconstruction. We have demonstrated the importance of automatically finding the right balance between structural and feature learning on a the diverse set of datasets with varying similarity between connected nodes, which are a reflection of real-world complex scenarios. Flex-GAD significantly reduces running time during training: (102x) per epoch compared to the Anomaly DAE and (3x) per epoch as compared to GAD-NR on average across 7 benchmarks. At the same time, when comparing with GAD-NR, the current SOTA GAD-GAE-based approach, our method achieved a 7.79\% improvement in average AUC.

\bibliographystyle{ACM-Reference-Format}
\bibliography{ref}


\end{document}